\documentstyle[prl,aps,multicol,epsf]{revtex}
\begin{document}
\draft

\title{Hall Coefficient and electron-electron interaction of 2D electrons in Si-MOSFET's.}

\author{Sergey~A.~Vitkalov}
\address{Physics Department, City College of the City
University of New York, New York, New York 10031} 
\date{\today}
\maketitle \begin{abstract}
Recent experiments in silicon MOSFETs indicate that the Hall coefficient is  
independent of magnetic field applied at a small angle with respect to the
plane. Below a scattering between spin-up and spin-down carriers 
is considered to be the main reason for the experimental observation. 
Comparison of two band model  with experiment provides an upper limit for the electron-electron
scattering time $\tau_{ee}$ in the  dilute 2D electron system as a function of
electron density $n_s$.  The time $\tau_{ee}$ increases gradually with $n_s$,
becoming much greater than the transport scattering time $\tau_p$ for densities $n_s>4
\times 10 ^{11}$ cm$^{-2}$.  Strong electron-electron scattering  is
found for $1.22 \times 10 ^{11} <n_s<3 \times 10 ^{11}$ cm$^{-2}$,  the region  
which is near to the apparent metal insulator transition.  
\end{abstract}

\pacs{PACS numbers: 72.15.Gd, 72.15.Lh, 72.25.Rb}

\begin{multicols}{2}

\section{Introduction}
 
Strongly interacting two-dimensional systems of electrons and holes have 
drawn intensive recent attention due to their anomalous behavior as a 
function of temperature and magnetic field\cite{rmp}: the resistance exhibits
metallic  temperature-dependence above a critical density, $n_c$, raising the
possibility of  a metallic phase  in two dimensions \cite{krav}.  An
additional  intriguing characteristic of these systems is their enormous
response to magnetic  fields applied in the 2D plane of the carriers
\cite{simonian,pudalov,cambridge,yoon}.  The resistivity increases
substantially with in-plane magnetic field and saturates to a new value
above a density-dependent characteristic magnetic field $H_{sat}$.  Several
experiments \cite{okamoto,vitkalov,Tutuc} have shown that the field $H_{sat}$ 
corresponds to the onset of full spin polarization of the 2D electron
system.  With increasing in-plane magnetic field the system thus evolves
from zero net spin  polarization, with equal numbers of spin-up and spin-down
electrons, to a completely spin-polarized state above the field $H_{sat}$. 

Recent measurements of the Hall resistance in parallel magnetic field 
\cite{vitkalov_hall} have revealed another unexpected physical property:
the Hall coefficient does not vary with parallel magnetic field for
fields ranging from $0$ to well above $H_{sat}$.  This is in apparent
contradiction with expectations based on straightforward arguments
\cite{Dolgopol2} that predict different mobilities of the spin-up and
spin-down electrons and, therefore, a substantial variation of the Hall
coefficient with in-plane magnetic field \cite{vitkalov_hall}.  The purpose
of the present paper is to provide a possible explanation of
the behavior of the Hall coefficient. The electron-electron ($e-e$)
scattering is considered as the main reason for the invariance of the Hall coefficient 
with in-plane magnetic field.  A comparison with experiment demonstrates that 
the frequency of the electron-electron scattering events $\nu_{ee}=1/\tau_{ee}$  
increases with decreasing electron density $n_s$.
At density $1.22\times 10 ^{11} <n_s<2 \times 10 ^{11}cm^{-2}$, the $e-e$ 
scattering rate $1/\tau_{ee}$  is found to be higher than the transport scattering
\vbox{
\vspace{-0.4 in}
\hbox{
\hspace{-0.3in} 
\epsfxsize 3.6 in \epsfbox{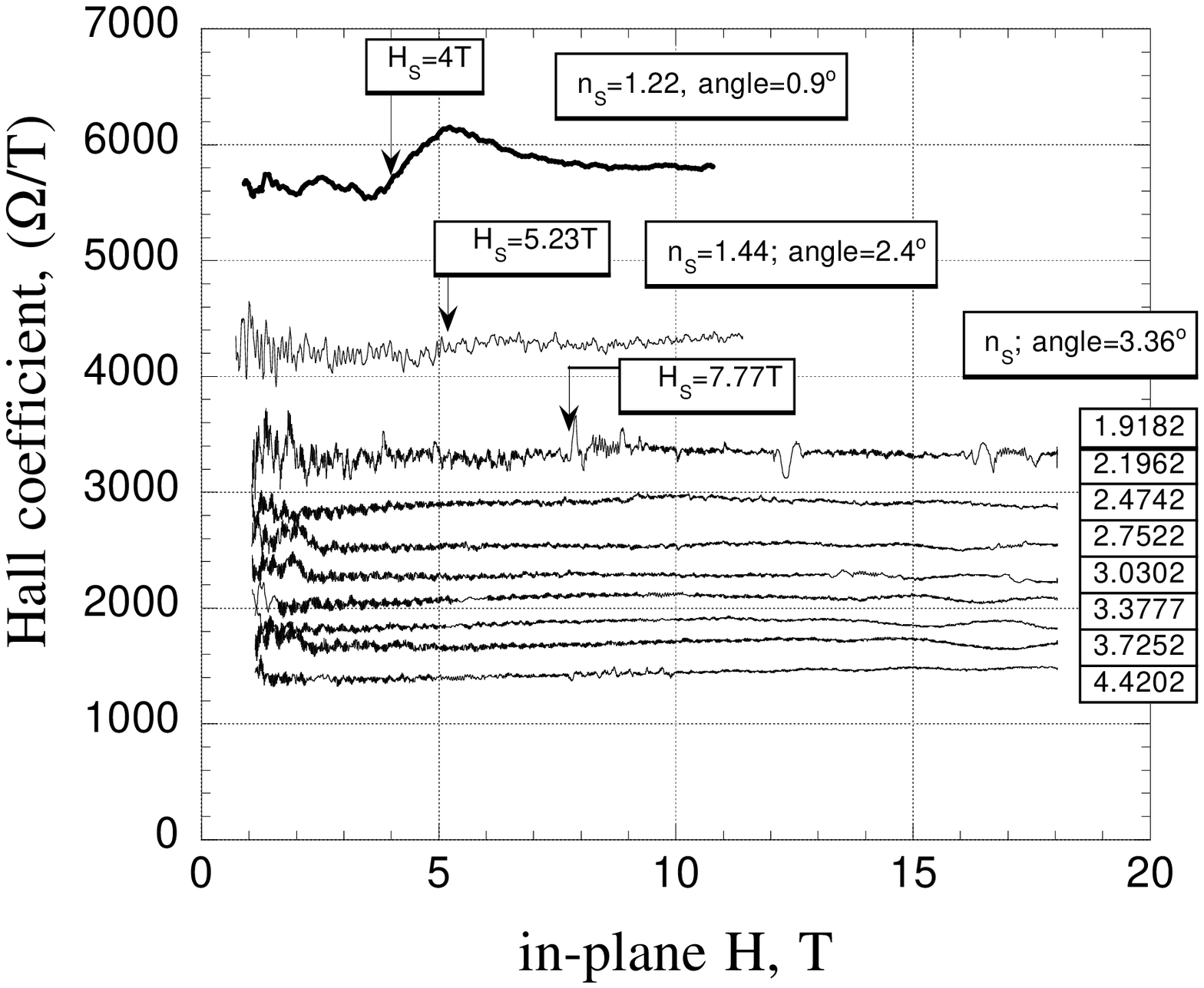} 
}
}
\vskip 0cm
\refstepcounter{figure}
\parbox[b]{3.3in}{\baselineskip=12pt FIG.~\thefigure.
The  Hall coefficient, $R_H = R_{xy}/H_\bot$, as a function of parallel magnetic 
field at different electron densities as labeled. Temperature is about 100mK for
 $n_s>1.9 \times 10^{11} cm^{-2}$ and is  250 mK for other curves. The arrows 
indicates the field $H_{s}$ of the complete spin polarization obtained from the 
saturation of the longitudinal magnetoconductivity\cite{ferro,accuracy}.   
\vspace{0.10in}
}
\label{1}  
rate $1/\tau_p$. This indicates that the $e-e$ interaction can be a dominant reason 
of decay of the electron states in the dilute 
2D system near the apparent metal-insulator  transition in the silicon MOSFET's 
at temperature above 0.1K.

\section{Experiment}
The data used in the paper were obtained  
on two silicon MOSFETs of comparable mobility $\mu \approx 20,000\;$V/(cm$^2s)$ 
at $T=4.2$ 
The samples were mounted on a rotating platform at the end of a low temperature 
probe.  Measurements were taken at temperature $T=0.235$~K and magnetic
fields $H$ up to $12$ T in a $^3$He Oxford Heliox system at City College of
NY. The Hall coefficient was measured at higher magnetic fields up to 18T at 
the National High Magnetic Field Laboratory in Tallahassee, Florida. 
The details of the experiment are presented in the paper\cite{vitkalov_hall,acknowlegment}. 
The Hall coefficient $R_H$, corresponding to different electron densities
from $n_s=1.22 \times 10^{11}$ cm$^{-2}$ to $n_s=4.42 \times 10^{11}$
cm$^{-2}$, is shown in Fig.1. $R_H$ is a weak function of the magnetic
field, varying by less the 3-7\% over a wide range of magnetic field for all
electron densities.  For several electron densities, the arrows mark the
magnetic field $H_s$ above which the longitudinal conductivity saturates \cite{ferro}
and 2D electrons are spin polarized completely \cite{accuracy}.
A considerable
change in $R_H$ is observed for the lowest electron density $n_s=1.22 \times
10^{11}$ cm$^{-2}$ at $H>4T$, where the system is in the insulating
regime\cite{kevin}. This behavior requires further investigation.  

The longitudinal conductivity $\sigma$ depends strongly on in-plane
magnetic field $H$\cite{ferro}.  
This indicates that the average mobility of the electrons
changes substantially with magnetic field. In the absence of interactions
between carriers, the mobility of spin-up and spin-down electrons should vary
independently of each other.  Taking into account the very different,
variable Fermi velocities of spin-up and spin-down electrons in an in-plane
field, we  expect that the mobilities of the two subbands are
most likely different in a magnetic field.  The Hall coefficient measurements
indicate, however, that the mobility of spin-up and spin-down electrons
in the field $H$ are  close to each other.  
Electron-electron interaction  are a promising candidate to explain this result. 
Indeed, it is known\cite{Gantmakher}, that due to the strong $e-e$ interband scattering  
the nonequilibrium averaged velocities of two different conducting bands  
in  an external electric field are, in fact, the same.  Below I will employ this 
approach for the 2D electron system, which is spin polarized by an external magnetic field.

\section{Model}  

Boltzman transport theory provides a description of the classical
magnetoconductivity and Hall resistance of carriers occupying several
different subbands in $k$-space\cite{Ziman,Gantmakher,Zaremba}.  For
example, the carriers in different subbands can have different electrical
charge (electrons and holes) or different kinematic properties such as mass
(light and heavy holes in GaAs\cite{Yaish}).  The 2D electrons in a Si-MOSFET
in a strong in-plane magnetic field occupy two subbands, one containing
spin-up and the other spin-down electron states.  Below I present a simple
description of a two band model, based on analysis of the total averaged
nonequilibrium momentum of each band in the mean free time approximation \cite{Ziman,Gantmakher}.
A more general approach can be found in reference \cite{Zaremba}.

In the absence of electron-electron scattering the spin-up and spin-down electrons can
be considered independently.  In this case we describe the relaxation of the
total momentum $\vec P_{\uparrow, \downarrow }$ of the nonequilibrium electron
distribution in each spin subband by a scattering time $\tau_{\uparrow,
\downarrow}$.  The electron-electron interaction (or scattering) induces a
redistribution of the non-equilibrium momentum of carriers between two spin
subbands and, therefore, changes the total momenta $\vec P_{\uparrow}$ and
$\vec P_{\downarrow }$. To account for the inter-subband redistribution of the nonequilibrium momentum 
  an effective force $< \vec F_{int}>$ acting between spin-up and
spin-down electron subbands is used\cite{Gantmakher,Vitkalov_drag}. 
The force $<\vec F_{int}>$ changes the total momentum of
each spin subband.  Newton's equations for total momentum $\vec P_{\uparrow}$,
($\vec P_{\downarrow}$) of the spin-up (spin-down) electrons in an external
electric field $\vec E$ and a normal magnetic field $\vec H_{\perp}$ yield:

$$
\frac{d \vec P_{\uparrow}}{dt}=en_{\uparrow}\vec E +\frac{e}{c}  n_{\uparrow} [\vec v_{\uparrow} 
\times \vec H_{\perp}]-\frac{\vec P_{\uparrow}}{\tau_{\uparrow}}+  <\vec F_{int}> \eqno{(1a)}$$

$$\frac{d \vec P_{\downarrow}}{dt}=en_{\downarrow}\vec E +\frac{e}{c}  n_{\downarrow} [\vec v_{\downarrow} 
\times \vec H_{\perp}]-\frac {\vec P_{\downarrow}}{\tau_{\downarrow}}-  <\vec
F_{int}>,
\eqno{(1b)}
$$ 
where $n_{\uparrow}, \vec v_{\uparrow}$, ($ n_{\downarrow}, \vec
v_{\downarrow} $) are the density and average velocity of spin-up (spin-down)
electrons. The functional dependence of the force $<\vec F_{int}>$ was
determined from the following observations \cite{Vitkalov_drag}.  We assumed that
each interaction (or collision) between two particles does not depend on 
other particles.  In this case the total effective  force $<\vec F_{int}>$
should be proportional to the product of the densities of spin-up and
spin-down electrons: 
$ <\vec F_{int}> \sim n_{\uparrow} \times n_{\downarrow} $. 
In the absence of a relative drift of the spin-up electrons with respect to
spin-down electrons ($\vec v_{\uparrow}= \vec v_{\downarrow} $) the  
force $ <\vec F_{int}>$ is expected to  be equal to 0.  This result is, in fact, a consequence
of invariance of the system under Galileo's transformations, if we neglect
variations of the electron distribution with drift velocity (in other words,
if we consider the linear response of the system).  In this approximation the
  force $ <\vec F_{int}>$ is proportional to the difference between
average velocities of spin-up and spin-down electrons: $ <\vec F_{int}>
\sim  (\vec v_{\uparrow}- \vec v_{\downarrow}) $. Thus  we use the following
expression for the  force $ <\vec F_{int}>$:
 
 $$
<\vec F_{int}>=\alpha n_{\uparrow}   n_{\downarrow}(\vec v_{\uparrow}- \vec
v_{\downarrow}),
\eqno{(2)}
$$
where $\alpha$ is a constant.  In stationary states the momenta $\vec
P_{\uparrow}, \vec P_{\downarrow}$ of the spin subbands are constant. 
Therefore the left sides of Eq.(1) are equal to $0$.  Combining Eqs. (1) and
(2) we obtain a linear relation between the electric field $\vec E$ and the
momenta  $\vec P_{\uparrow}, \vec P_{\downarrow}$ of the spin subbands:

$$
(i\omega_c+\epsilon_{\uparrow})  P_{\uparrow}- \gamma_{\uparrow}P_{\downarrow}=en_{\uparrow} E \eqno{(3a)}$$
$$(i\omega_c+\epsilon_{\downarrow})  P_{\downarrow}-
\gamma_{\downarrow}P_{\uparrow}=en_{\downarrow} E, 
\eqno{(3b)}
$$
where $P=P_x+iP_y$, $E=E_x+iE_y$, 
$\gamma_{ \uparrow, \downarrow }=\alpha n_{\uparrow, \downarrow }/m$, 
$\epsilon_{ \uparrow, \downarrow }=1/\tau_{\uparrow, \downarrow }+  \gamma_{\downarrow ,
\uparrow }$, $\omega_c=eH_{\perp}/mc$ is the cyclotron frequency  and $m$ is the band mass 
of the electrons. The current density is given by

$$
J= en_{\uparrow}v_{\uparrow}+ en_{\downarrow}v_{\downarrow} =
\frac{e}{m}(P_{ \uparrow} + P_{ \downarrow})
\eqno{(4)}
$$   

The solution of the linear equations (3) gives the subband momenta $P_{
\uparrow}$ and $P_{ \downarrow}$ as a function of external electric and
magnetic fields.  The longitudinal and Hall conductivities of the system
were obtained using the real  and imaginary parts of Eq.(4), respectively:

$$
\sigma_{xx}=\frac {(\epsilon_{\uparrow} \epsilon_{\downarrow} -\gamma_{\uparrow} 
\gamma_{\downarrow} -\omega_c^2   )[n_{\uparrow}(\epsilon_{\downarrow}+\gamma_{\downarrow})+
n_{\downarrow }(\epsilon_{\uparrow }+\gamma_{\uparrow } )]+}
{(\epsilon_{\uparrow} \epsilon_{\downarrow}-\gamma_{\uparrow} \gamma_{\downarrow} -
\omega_c^2)^2+ \omega_c^2(\epsilon_{\uparrow }+\epsilon_{\downarrow })^2} 
$$
$$
\frac{ +\omega_c^2(\epsilon_{\uparrow }
+\epsilon_{\downarrow })(n_{\uparrow }+n_{\downarrow })} { } 
\eqno{(5)}
$$

$$
\sigma_{xy}=\frac { \omega_c [(n_{\uparrow}(\epsilon_{\downarrow}+\gamma_{\downarrow})+
n_{\downarrow }(\epsilon_{\uparrow }+\gamma_{\uparrow } ))(\epsilon_{\uparrow }+
\epsilon_{\downarrow }) -}
{(\epsilon_{\uparrow} \epsilon_{\downarrow}-\gamma_{\uparrow} \gamma_{\downarrow}
 -\omega_c^2)^2+ \omega_c^2(\epsilon_{\uparrow }+\epsilon_{\downarrow })^2} 
$$
$$
  \frac {-(n_{\uparrow }+n_{\downarrow })(\epsilon_{\uparrow} 
\epsilon_{\downarrow}-\gamma_{\uparrow} \gamma_{\downarrow} -\omega_c^2) } { }
\eqno{(6)}  
$$ 
The Hall resistivity was obtained by inverting the conductivity tensor
$\sigma$.

In order to evaluate the strength of the electron-electron scattering 
quantitatively we 
consider a model for the in-plane field dependence of the longitudinal
conductivity $\sigma(H)$.  One possible explanation of the field dependence
of $\sigma(H)$ was proposed recently by Dolgopolov and Gold (DG)
\cite{Dolgopol2}.  They have argued that the screening of electrically
charged impurities by the 2D electrons depends substantially on the
populations of the spin-up and spin-down subbands and therefore varies
with in-plane magnetic field.  This causes the resistance to increase with
field $H$, saturating when the electrons reach full spin polarization at
$H>H_{s}$.  In the DG approach \cite{Dolgopol2} the average scattering
probabilities, $1/\tau_{\uparrow,\downarrow}$, are calculated corresponding
to spin-up and  spin-down electrons in a parallel magnetic field.  The
longitudinal conductivity is  the sum of contributions of spin-up and spin-down bands: 
$\sigma=\sigma^\uparrow + 
\sigma^\downarrow=e^2n_{\uparrow } \tau_{\uparrow}/m + e^2n_{\downarrow} \tau_{\downarrow}/m$, where 
$n_{\uparrow,\downarrow}=n_0(1 \pm \xi)$  is the density of the spin-up 
(spin-down) electrons and $\xi=H/H_{s}, (H<H_{s})$ is the degree of spin 
polarization of the 2D system.  We use their expression for the conductivity 
$\sigma(H)$ \cite{Dolgopol2} to find the scattering times $\tau_{\uparrow \downarrow}(H)$
 as a function of parallel magnetic field $H$. 
In accordance with the eq.(5,6) the variations of the Hall coefficient $R_H(H)$ are  result 
of different mobilities of spin-up and spin-down electrons in a magnetic field H. 
Therefore we expect a similar  dependence of the $R_H(H)$ (and the $e-e$ scattering rate) 
for any  other models of magnetoconductivity corresponding to the experiment, in which 
the bare mobilities of spin-up and spin-down electrons are different considerably.

\vbox{
\vspace{0 in}
\hbox{
\hspace{-0.3in} 
\epsfxsize 4 in \epsfbox{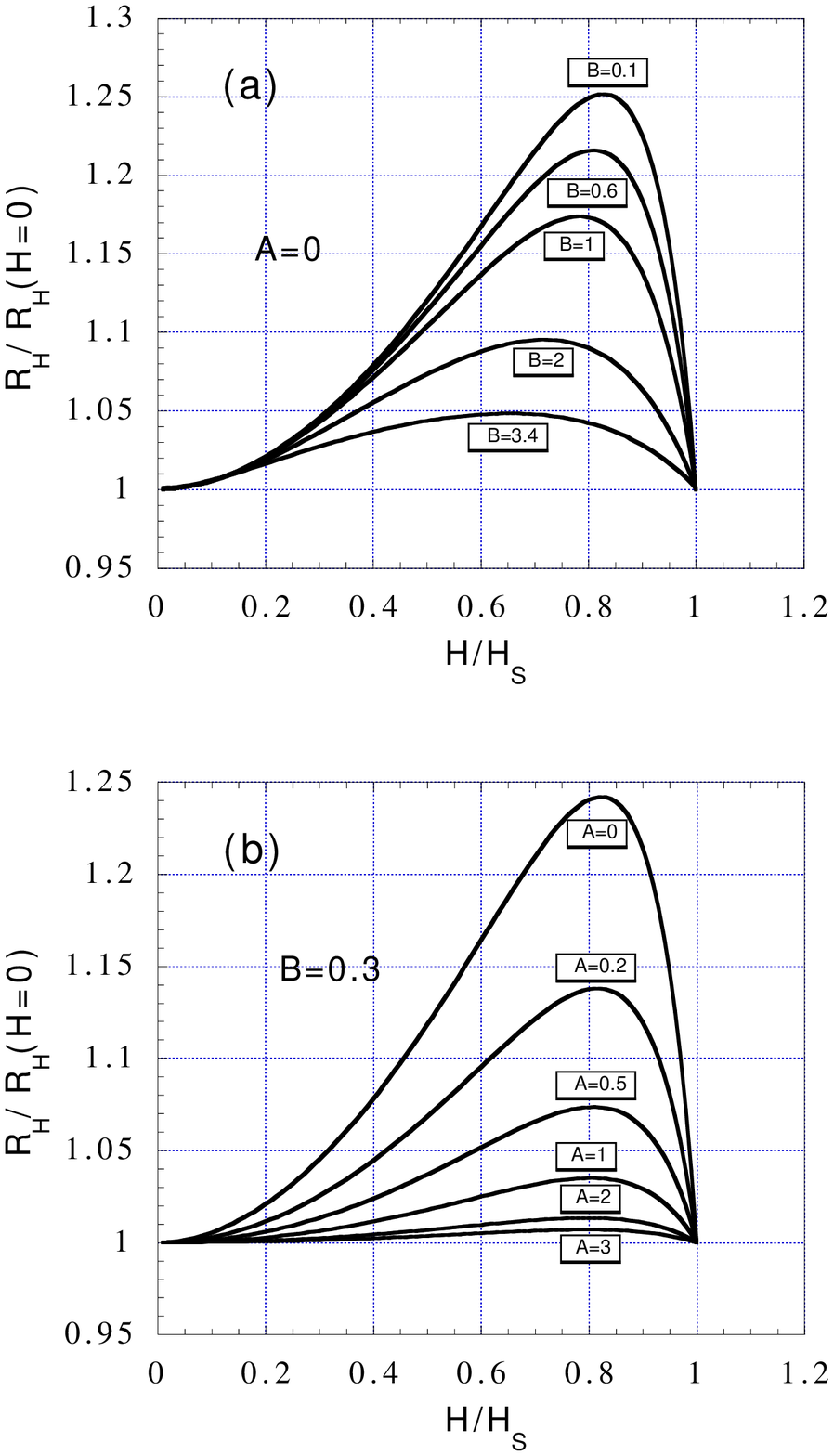} 
}
}
\refstepcounter{figure}
\parbox[b]{3.3in}{\baselineskip=12pt FIG.~\thefigure.
(a) The Hall coefficient as a function of the spin polarization $\xi=H/H_s$ at different 
value of the parameter $B =(\omega_c \tau_0)_{H=H_s}$ as labeled in the absent of the $e-e$ scattering
($A=0$).  (b) The Hall coefficient as a function of the spin polarization
 $\xi=H/H_s$ at different value of the $e-e$ scattering $A= \tau_0/\tau_{ee}$ as 
labeled.  The parameter $B =0.3$   
\vspace{0.10in}
}
\label{2}

According to Eq.(5,6), the behavior of the Hall coefficient
as function of $H$ depends on two parameters.  One is the strength of 
\vbox{
\vspace{-0.2in}
\hbox{
\hspace{-0.4in} 
\epsfxsize4in \epsfbox{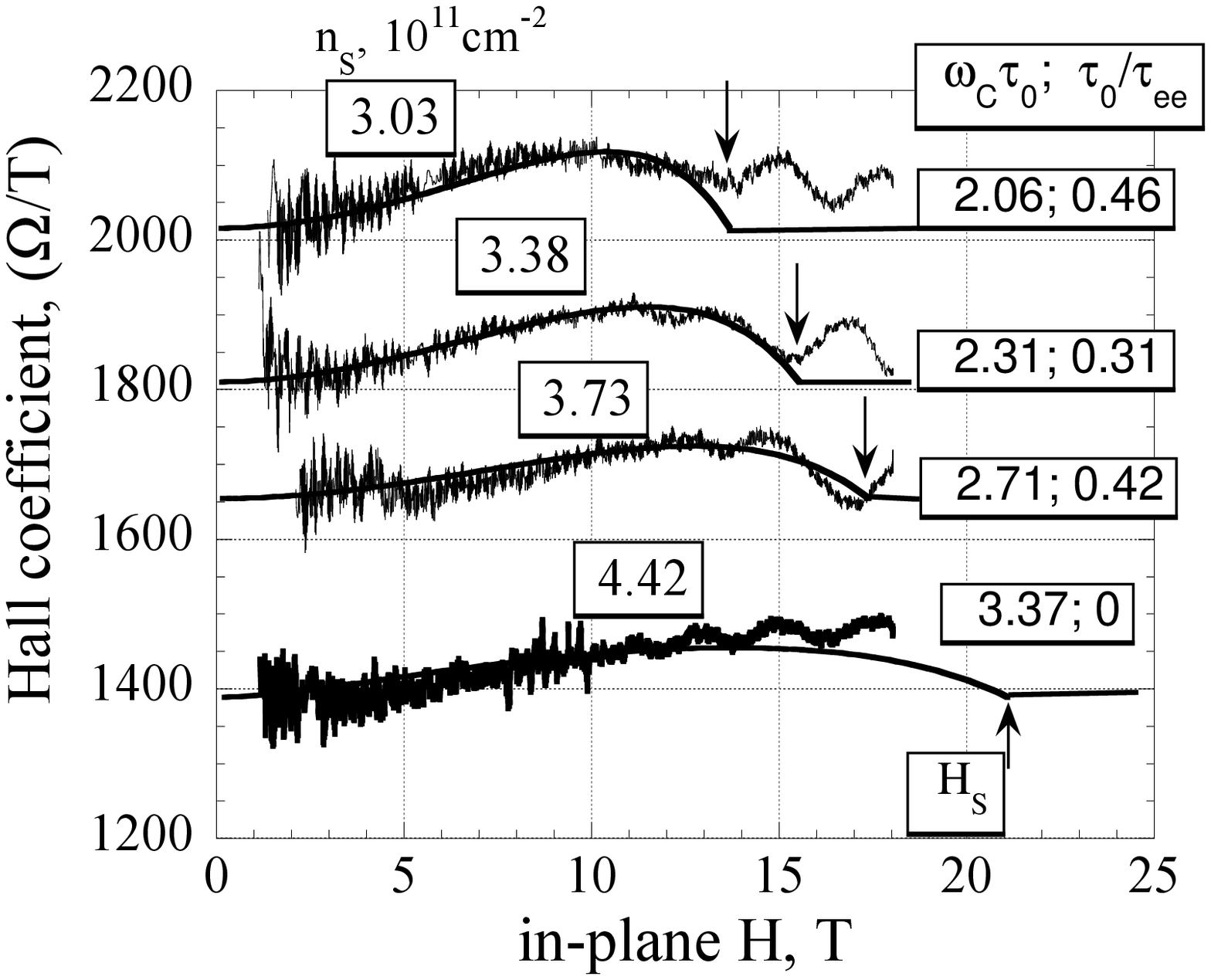} 
}
}
\refstepcounter{figure}
\parbox[b]{3.3in}{\baselineskip=12pt FIG.~\thefigure.
 The  Hall coefficient as  a function of the in-plane magnetic field at high electron density.
 The solid lines are numerical fit corresponding to the two band  model (see eq. (5,6)). 
The parameter $B =(\omega_c \tau_0)_{H=H_s}$ was obtain as a ratio of the $R_{xy}(H=H_s)/\rho(H=0)$ 
for each curve. The spin polarization field $H_s$ was found from the saturation of 
the magnetoconductivity\cite{ferro}. The arrows indicate the field $H_s$ for each electron density.
The ratio of the transport scattering time $\tau_p$ to $e-e$ scattering time $\tau_{ee}$, 
founded from the fitting, are labeled.  
\vspace{0.10in}
}
\label{3}
the $e-e$ interaction, which we characterize by a factor $A=\alpha n_0 \tau_0/m=\tau_0/\tau_{ee}$, the ratio
of the inter subband $e-e$ scattering rate to the transport scattering rate at $H=0$T. The second
parameter is $B =(\omega_c \tau_0)_{H=H_s}$ calculated at field $H=H_s$, 
where the time $\tau_0$ is the scattering time at $H=0$ T without $e-e$ interaction ($A=0$).  
In Fig. (2a) we
show the Hall coefficient calculated for different scattering times $\tau_0$
(B) without $e-e$ scattering ($A=0$). The Hall coefficient depends
substantially on the parameter $B$.  The behavior of the Hall coefficient for
different $B$ and $A=0$ is similar to results obtained earlier, neglecting the 
intersubband $e-e$ scattering\cite{Herbut,note}. 
In the experiment, the parameter $B$ was varied
between $0.1$ and $3.4$ depending on electron density and the angle $\phi$
between the magnetic field and the plane of the electrons. In accordance with
Fig.2a for all angles and electron densities measured excepting the highest,
we should expect a substantial change of the Hall coefficient in the absence
of $e-e$ scattering ($A=0$).  In Fig.2b we present the Hall coefficient as a
function of in-plane magnetic field at $B=0.3$ and different rate of $e-e$ scattering. 
 The Hall coefficient behavior depends substantially on the $e-e$ interband
scattering.  For $A=1$ the variation of the Hall coefficient is less than
4\%.
\vbox{
\vspace{0in}
\hbox{
\hspace{-0.2in} 
\epsfxsize 3.6in \epsfbox{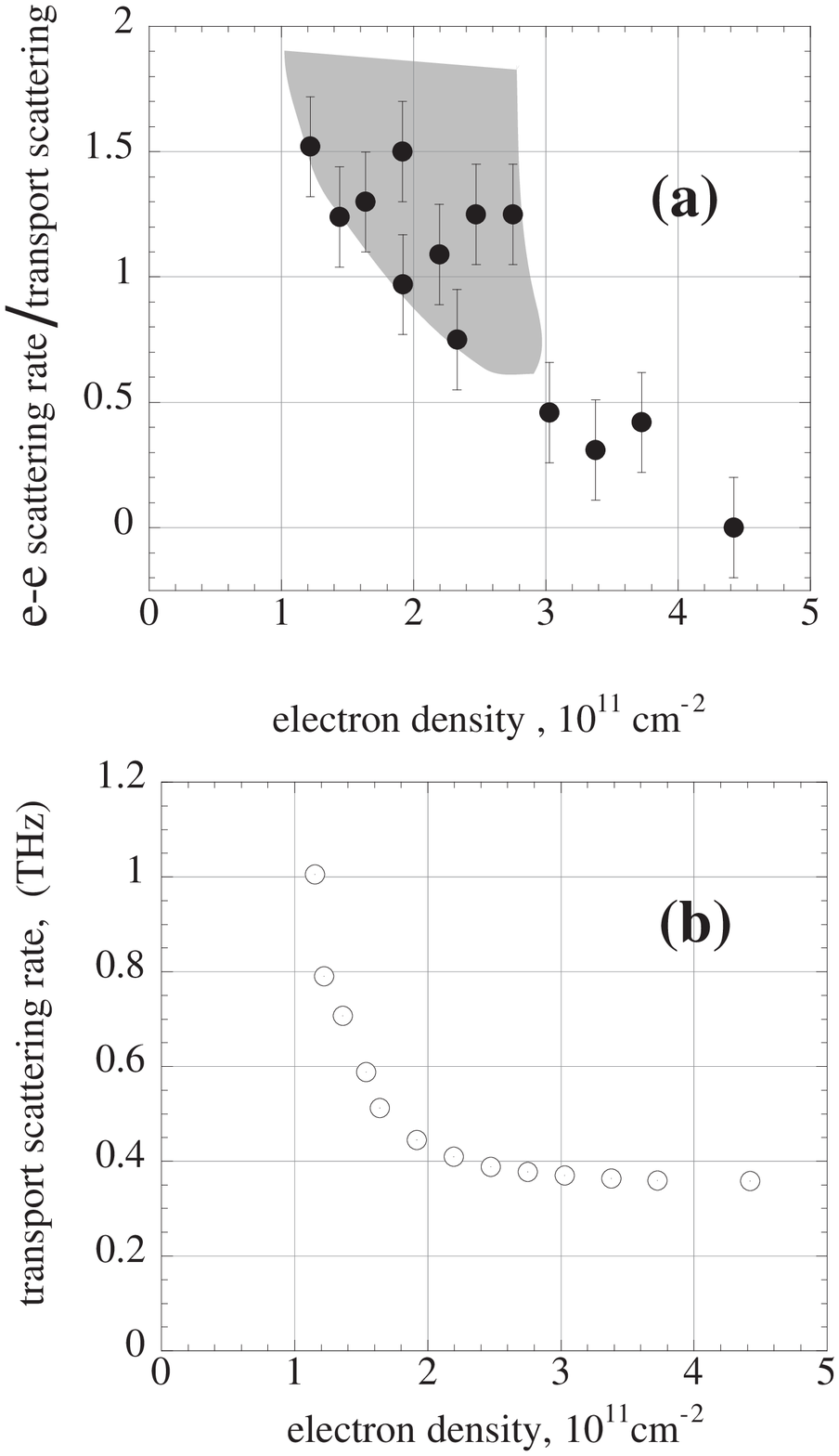} 
}
}
\vskip 0.5cm
\refstepcounter{figure}
\parbox[b]{3.3in}{\baselineskip=12pt FIG.~\thefigure.
(a) The ratio of the electron-electron scattering rate $1/\tau_{ee}$ to the rate of the 
transport scattering $1/\tau_p$ at different electron densities $n_s$. The shadow area indicates
 the possible values of the $e-e$ scattering rate in the dilute 2D electron system. 
(b) the transport scattering rate $1/\tau_p$ vs $n_s$.   
\vspace{0.10in}
}
\label{4}    
\section{Comparison with experiment}
We now compare the results of the two band model presented above with 
experiment.  The parameter $B =(\omega_c \tau_0)_{H=H_{s}}$ was found from
experiment as the ratio of the Hall resistance
$\rho_{xy}=H_{\perp}/(nec)(H=H_{s})$ at magnetic field $H=H_{s}$ to the 
longitudinal resistivity $\rho_{xx}=m/(ne^2 \tau_0)$ at $H=O$T. The  strength
of the $e-e$ scattering (coefficient $A$) was varied to obtain agreement with
experiment.  A comparison is shown on Fig.3 for four different electron
densities.  The accuracy of the experiment is not sufficient to extract the 
value of the $e-e$ scattering time $\tau_{ee}$ in the system at electron
densities below $n_s=3 \times 10^{11}$ cm$^{-2}$.  The main difficulty is 
that the   time $\tau_{ee}$   appears to be too short and comparable
with the  mean free time of the electrons $\tau_p$ at $n_s<3 \times 10^{11}$
cm$^{-2}$.  We determined an upper limit for $\tau_{ee}$ or lower limit for
the spin scattering rate $\nu_{ee}$ for electron densities $n_s<3 \times 10^{11}$
cm$^{-2}$.  The ratio of the electron-electron scattering  rate to the transport scattering rate
extracted from the experiment is presented in Fig.4.  The transport scattering rate $1/\tau_p$, 
obtained from the Drude conductivity ( neglecting the $e-e$ mass enhancement), is
shown for comparison.  Fig.4 demonstrates that the electron-electron scattering is  
strong in the dilute 2D system of electrons in SI-MOSFETs at electron density 
below $n_s=3 \times 10^{11}cm^{-1}$.  It increases with
decreasing electron density and becomes higher than the transport scattering
rate at $n_s<2 \times 10^{11}$ cm$^{-2}$.
  
In summary, we have shown that the independence of the Hall coefficient of 
dilute 2D electrons in SI-MOSFETs on in-plane magnetic field can be the result of 
electron-electron scattering between two spin subbands.  
Comparison of experimental results with a two band model yields 
  the electron-electron scattering rate 
$\nu_{ee}=1/\tau_{ee}$ between spinup and spindown 2D electrons in Si-MOSFETs. 
The $e-e$ scattering was found to be a decreasing function of the electron density  
at $1.22 \times 10^{11} <n_s<4\times 10^{11}$ cm$^{-2}$ in accordance with general 
expectations.  At $1.22 \times 10^{11} <n_s<3
\times 10^{11}$ cm$^{-2}$ the $e-e$ scattering rate $\nu_{ee}$ of the 2D electrons
 in silicon MOSFET's is comparable or
even higher that the transport scattering rate $1/\tau_p$ at temperature $T>0.1K$ .

\section{ACKNOWLEDGMENTS}

I thank V.T. Dolgopolov, A. I. Larkin and V. Falko for helpfull discussions, 
M. P. Sarachik for support  and comments of the manuscript, and T. M.  Klapwijk 
for providing of the MOSFET's. 
This work was supported by DOE grant No. DOE-FG02-84-ER45153. 
Partial support was also provided by NSF grant DMR-9803440.

\end{multicols} 


\begin{references}

\bibitem{rmp} E. Abrahams, S. V. Kravchenko, and M. P. Sarachik, preprint
cond-mat/0006055 (2000) (to be published in Rev. Mod. Phys.); M. P. Sarachik and 
S. V. Kravchenko, Proc. Natl. Acad. Sci. {\bf 96}, 5900 (1999).
\bibitem{krav} S.\ V.\ Kravchenko, G.\ V.\ Kravchenko, J.\ E.\
Furneaux, V.\ M.\ Pudalov, and M.\ D'Iorio, Phys.\ Rev.\ B {\bf 50}, 8039
(1994); S.\ V.\ Kravchenko, W.\ E.\ Mason, G.\ E.\ Bowker,
J.\ E.\ Furneaux, V.\ M.\ Pudalov, and M.\ D'Iorio, Phys.\ Rev.\ B {\bf 51},
7038 (1995); S.\ V.\ Kravchenko, D.\ Simonian, M.\ P.\ Sarachik, W.\ E.\ Mason, and J.\ E.\ Furneaux, Phys.\
Rev.\ Lett. {\bf 77}, 4938 (1996).
\bibitem{simonian} D.~Simonian, S.~V.~Kravchenko, M.~P.~Sarachik, and
V.~M.~Pudalov, Phys.\ Rev.\ Lett. {\bf 79}, 2304 (1997).
\bibitem{pudalov} V.~M.~Pudalov, G.~Brunthaler, A.~Prinz, and G.~Bauer,
Pisma Zh. Eksp. Teor. Fiz. {\bf 65}, 887 (1997)
[JETP Lett. {\bf 65}, 932 (1997)].
\bibitem{cambridge} M.~Y.~Simmons, A. R. Hamilton, M. Pepper, E. H.
Linfield, P. D. Rose,
D. A. Ritchie, A. K. Savchenko, and T. G. Griffiths, Phys.\ Rev.\ Lett. {\bf
80}, 1292 (1998).
\bibitem{yoon} J. Yoon, C. C. Li, D. Shahar, D. C. Tsui, and M. Shayegan,
Phys.\ Rev.\ Lett. {\bf 84}, 4421 (2000).
\bibitem{okamoto} T. Okamoto, K. Hosoya, S. Kawaji, and A. Yagi, Phys. Rev.Lett. 
{\bf 82}, 3875 (1999).
\bibitem{vitkalov} S. A. Vitkalov, H. Zheng, K. M. Mertes, M. P. Sarachik, and 
T. M. Klapwijk, Phys.\ Rev.\ Lett. {\bf 85}, 2164 (2000).
\bibitem{Tutuc}  E. Tutuc, E.P. De Poortere, S.J. Papadakis, M. Shayegan 
 preprint cond-mat/0012128 (2000).
\bibitem{vitkalov_hall}  S. A. Vitkalov, H. Zheng, K. M. Mertes, M.
P. Sarachik, and T. M. Klapwijk, Phys. Rev. {\bf B}, to be published
\bibitem{Dolgopol2} V. T. Dolgopolov and A. Gold, JETP Letters, {\bf 71},27,(2000).
\bibitem{acknowlegment} I thank M. P. Sarachik and T. M. Klapwijk for putting  the experimental data 
at my disposal.
\bibitem{ferro} S. A. Vitkalov, H.Zheng, K. M. Mertes, M. P. Sarachik, and T. M. Klapwijk, 
preprint cond-mat/0009454 (2000).
\bibitem{accuracy} The saturation field $H_{sat}$\cite{ferro} and the field of complete spin 
polarization $H_s$ correspond to each other with an estimated accuracy about 1T at 
$n_s>1.6 \times 10^{11} cm^{-2}$ \cite{vitkalov,SdHrotation}.
\bibitem{SdHrotation}  S. A. Vitkalov, M.P. Sarachik, and T. M. Klapwijk, preprint cond-mat/0101196 (2000).
\bibitem{kevin}  K. M. Mertes, H. Zheng, S. A. Vitkalov, M. P. Sarachik, and 
T. M. Klapwijk, Phys. Rev. {\bf B}, Rapid Comm., {\bf 63}, 041101(R)  (2001). 
\bibitem{Ziman} J. M. Ziman, "Principles of the Theory of Solids", Cambridge 
University Press, (1972).
\bibitem{Gantmakher} V.F. Gantmakher and Y.B. Levinson, Sov. Phys. JETP {\bf 47}, 133 (1978).
\bibitem{Zaremba} E. Zaremba, Phys. Rev.B {\bf 45}, 14143 (1992). 
\bibitem{Yaish} Yuval Yaish, Oleg Prus, Evgeny Buchstab, Shye Shapira, Gidi Ben Yoseph, Uri Sivan, and Ady Stern Phys. Rev. Lett. {\bf 84},4954 (2000)
\bibitem{Vitkalov_drag} S.A. Vitkalov JETP Lett.{\bf 67}, 276 (1998).
\bibitem{Herbut}Igor F. Herbut preprint, cond-mat/0009164 (2000)
\bibitem{note} Substantial overestimation 
of the experimental scattering time $\tau_0$ by factor 10 was done in the paper \cite{Herbut}. 
As result a wrong conclusion was drawn regarding the experiment \cite{vitkalov_hall}.  
 

 
\end{references}
\end {document}